\documentclass{PoS}

\usepackage{bm}

\newcommand{\be}{\begin{equation}}
\newcommand{\ee}{\end{equation}}
\newcommand{\ba}{\begin{eqnarray}}
\newcommand{\ea}{\end{eqnarray}}
\newcommand{\bi}{\begin{itemize}}
\newcommand{\ei}{\end{itemize}}

\renewcommand{\>}{\rangle}

\newcommand{\fig}{Fig.~}

\newcommand{\la}{\label}

\newcommand{\ud}{\,\mathrm{d}}

\newcommand{\bx}{\boldsymbol{x}}

\title{Vector correlator and scale determination \\ in lattice QCD}

\ShortTitle{Vector correlator and scale determination in lattice QCD}

\author{Anthony Francis, Georg von Hippel, \speaker{Harvey B.\ Meyer}  
\\
        PRISMA Cluster of Excellence,
Institut f\"ur Kernphysik and Helmholtz~Institut~Mainz,
Johannes~Gutenberg-Universit\"at~Mainz,
D-55099 Mainz, Germany\\
        E-mail: \email{meyerh@kph.uni-mainz.de}}

\author{Fred Jegerlehner\\
        Humboldt-Universit\"at zu Berlin, Institut f\"ur Physik, Newtonstrasse 15, D-12489 Berlin;\\
       Deutsches Elektronen-Synchrotron (DESY), Platanenallee 6, D-15738 Zeuthen, Germany}

\abstract{We implement a proposal made in~\cite{Bernecker:2011gh} to determine the
  lattice spacing by matching the lattice vector correlator at a
  reference distance scale with the same correlator obtained by a
  dispersion relation based on the $R$-ratio determined
  experimentally. We work with the isovector current, requiring a
  separation of the isovector hadronic final states on the
  phenomenological side. We also discuss the finite-size effect on the
  correlator, which must be controlled in order for the method to be
  applicable.  }

\FullConference{31st International Symposium on Lattice Field Theory - LATTICE 2013\\
		July 29 - August 3, 2013\\
		Mainz, Germany}

\begin{document}

\section{Introduction}

Determining the lattice spacing in physical units is a central task in
almost any lattice QCD calculation. Pragmatically, the question is,
which dimensionful but renormalized quantity can be determined with
the highest overall accuracy on the lattice, statistical and
systematic errors both being taken into account. \emph{Relative scale
  setting} does not require the quantity to be known experimentally,
and is sufficient to compare all other dimensionful quantities
across different lattice spacings.  Quite a few choices of this
kind are available: the length scales $r_0$~\cite{Sommer:1993ce},
$r_1$~\cite{Bernard:2000gd} are based on the static potential. The
renormalized eigenvalue density of the Dirac operator $\nu_{\rm R}/V$
\cite{Giusti:2008vb} is also an attractive quantity for this purpose,
and so are $t_0$~\cite{Luscher:2010iy} and
$w_0$~\cite{Borsanyi:2012zs}, which are derived from the Wilson flow.

For the purpose of \emph{absolute scale setting} however, the quantity
must be known experimentally to a high degree of precision, both in
terms of statistical and systematic precision.  Commonly used
quantities are spectroscopic quantities such as the Omega
mass~\cite{Allton:2008pn}, and the decay constants of the light
pseudoscalar mesons, $F_\pi$ (see \cite{Durr:2013goa} for a recent
calculation of this quantity) and $F_K$~\cite{Fritzsch:2012wq}.  Note
also that at a time when most lattice calculations were done in the
quenched approximation, $r_0$ was used for absolute scale setting.
What quantities are considered to be known in a model-independent way
on the phenomenological side thus depends to some extent on the
targeted accuracy.

Here we explore a proposal to set the absolute scale using the vector
correlator~\cite{Bernecker:2011gh}. The point of view taken here is
that for the time being, phenomenology is more accurate than lattice
QCD in calculating this correlator in the Euclidean domain, and that
it is therefore reasonable to set the scale in this way. One immediate
advantage of this method is that a whole \emph{function} can be
compared between the lattice calculation and the phenomenological
extraction from the reaction $e^+e^-\to{\rm hadrons}$ and $\tau$
decays via a dispersion relation. The details of the scale-setting
condition can thus be optimized for the needs of lattice calculations.

\section{Scale setting via the isovector vector correlator}

Let $j_\mu^{\rm em}(x) $ be the electromagnetic current of hadrons.
Its Euclidean correlator,
\[
G^{\rm em}(t)\equiv  \int \ud^3\bx\; \<j^{\rm em}_z(t,\bx) j^{\rm em\,\dagger}_z(0)\>,
\]
admits the spectral representation:
\ba
G^{\rm em}(t) &=&  \int_0^\infty \ud\omega \, \omega^2\rho(\omega^2) e^{-\omega |t|},
\nonumber \\
\rho(s) &=& \frac{R(s)}{12\pi^2}, \qquad  R(s) \equiv  \frac{\sigma(e^+e^-\to {\rm hadrons})}
 {4\pi \alpha(s)^2 / (3s) } .
\nonumber
\ea
These equations provide the connection between lattice QCD observables and 
phenomenology.
If all exclusive channels are measured on the experimental side, the isoscalar/isovector
flavor separation can be done in a model independent way (assuming isospin symmetry).
We therefore introduce $R_1(s)$, defined as $R(s)$, but where the final hadronic state 
is required to be isovector. 

On the lattice, we need only compute the Wick-connected diagram for
the isovector correlator $G(t)$, and will therefore focus on this
case. Since we will use $N_{\rm f}=2$ ensembles, the isovector vector
current cannot create a hadronic state with hidden strangeness
(e.g.\ $K\bar K$ pairs), and therefore final states containing kaons
were removed from the spectral density on the phenomenological
side. Needless to say, the absence of strange quarks in our
simulations leads to a systematic uncertainty in the comparison of
lattice and phenomenological results.

Among many possible choices~\cite{Bernecker:2011gh}, one convenient
definition to set the scale based on the vector correlator is
\be\la{eq:325}
f(\tau_1) \equiv 3.25, \qquad f(t) \equiv t \cdot m_{\rm eff}(t) \equiv - \frac{t}{G(t)} \frac{dG}{dt}.
\ee
Any multiplicative renormalization factor on the vector current obviously cancels out in this quantity.
Summing up the contribution of the exclusive $I=1$ hadronic states in $e^+e^-$ data up to about 2\,GeV 
yields\footnote{A detailed description of this analysis will be published in a forthcoming paper.}
\be\la{eq:tau1pheno}
\tau_1 = 0.727(9){\rm \,fm}.
\ee
The region above 2\,GeV makes an almost negligible contribution at Euclidean time $\tau_1$
and can safely be treated using perturbation theory.
The specific choice of 3.25 is motivated by an approximate minimization of the relative 
error on $\tau_1$.
The propagation of the statistical error on the effective mass onto $\tau_1$ is given by 
\be
\frac{\delta \tau_1}{\tau_1} = 
\left( \frac{\tau_1}{f}\frac{d f}{ d\tau_1}\right)^{-1} \frac{\delta m_{\rm eff}(\tau_1)}{m_{\rm eff}(\tau_1)}
\stackrel{f(\tau_1)=3.25}{\approx } 2.2  \;\frac{\delta m_{\rm eff}(\tau_1)}{m_{\rm eff}(\tau_1)}, 
\ee
where the estimate of the slope is taken from the phenomenological
curve. Thus for physical pion masses, a one percent error on the
effective mass translates into a 2.2 percent error on the scale
determination.  This loss of precision is due to the relatively flat
behavior of the function $f(t)$. Clearly, $f(t)$ would be a constant
in a scale-invariant theory, in particular it would be 3 for
non-interacting massless quarks.  This approximately $t$ independent
behavior `accidentally' extends far beyond the region of validity of
perturbation theory due to a shallow minimum around $t=0.5{\rm \,fm}$.
A preliminary study shows that this undesirable feature is largely
absent in the $N_{\rm f}=2+1$ flavor theory. For large (but not
asymptotic) $t$, $\frac{t}{f(t)} \frac{df}{dt} \approx 1$ since $f(t)
\approx m_\rho t$, but then the relative error on the effective mass
tends to be significantly larger.

\begin{table}
\centerline{\begin{tabular}{c@{~~~~~}c@{~~~~~}c@{~~~~~}c} 
\hline
$\beta$ &  label & $L/a$ & $m_\pi[{\rm MeV}]$ \\
\hline
5.2 & $A_4$ & $32$   &  380 \\
    & $A_5$ & $32$    & 330 \\
\hline 
5.3 & $F_6$ & $48$  & 310 \\ 
    & $F_7$ & $48$   & 270 \\
    & $G_8$ & $64$  & 190 \\
\hline
5.5 & $N_5$ & $48$  &  440 \\
    & $N_6$ & $48$  & 340 \\
    & $O_7$ & $64$ & 270 \\
\hline
\end{tabular}}
\caption{List of the CLS ensembles used, with the linear spatial extent $L$ and the pion masses. 
All ensembles have a time extent $T=2L$ and are such that $m_\pi L> 4$.}
\la{tab:list}
\end{table}

\begin{figure} 
\centerline{\includegraphics[width=.55\textwidth]{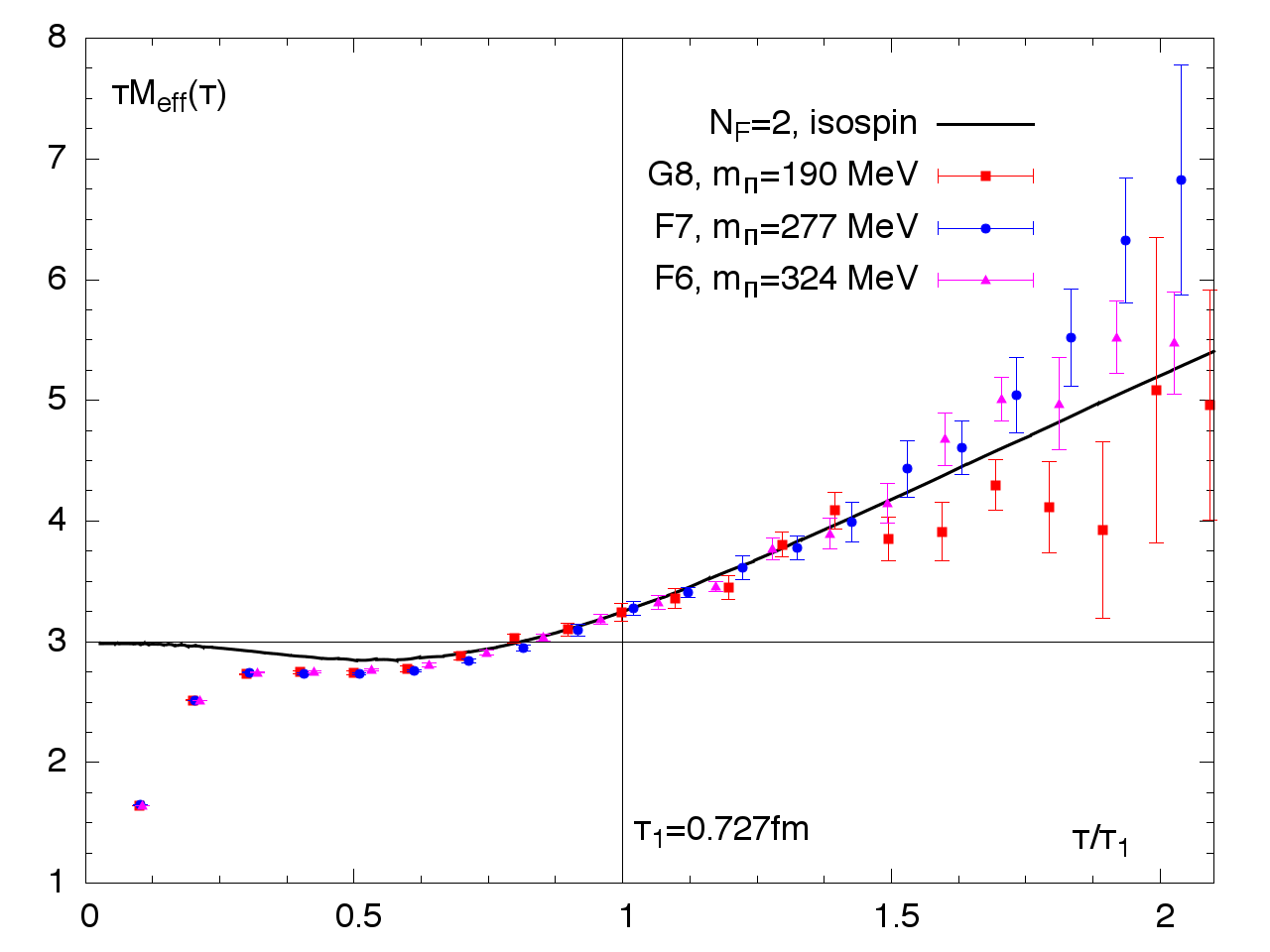} 
\includegraphics[width=.55\textwidth]{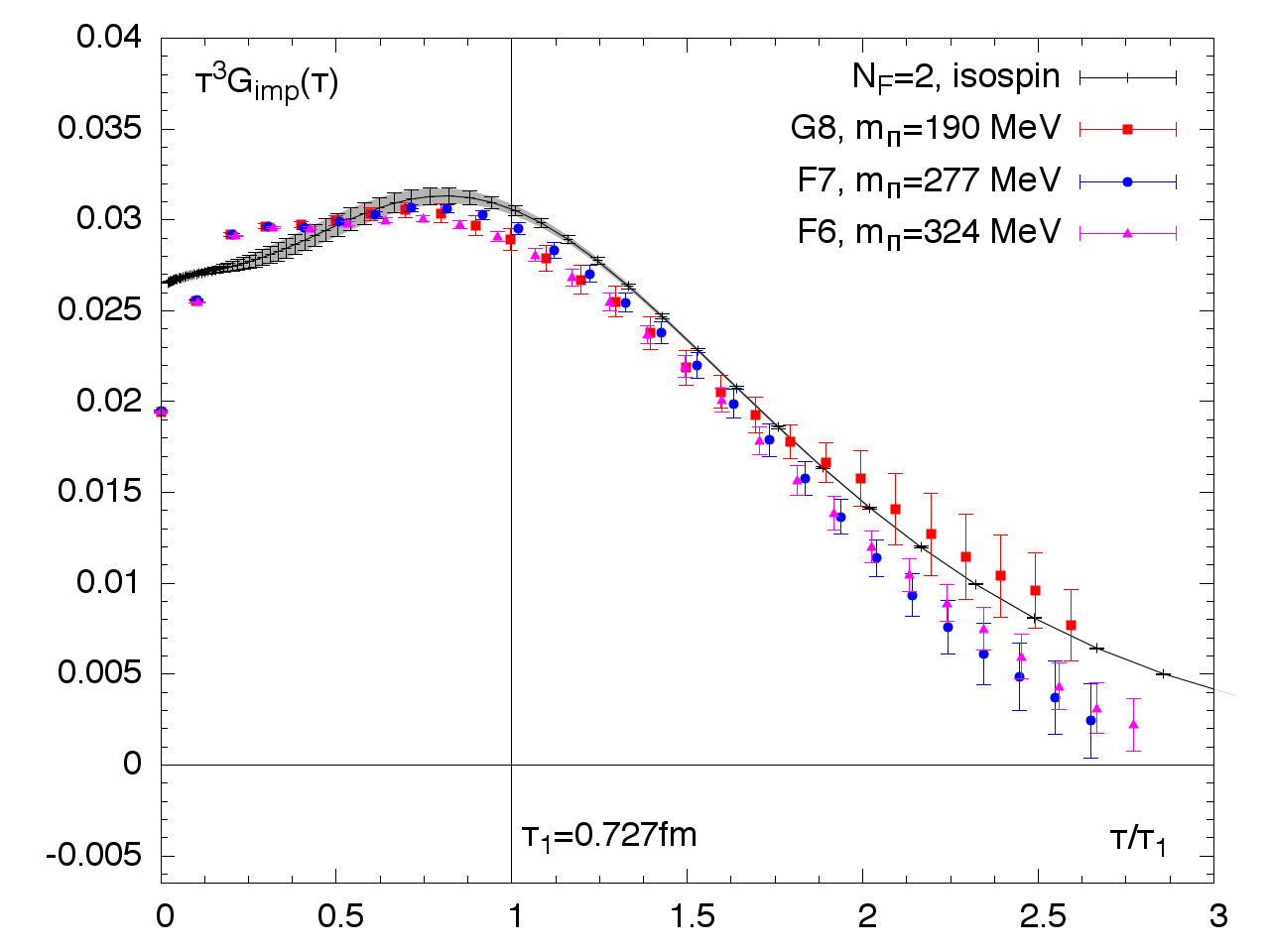}}
\caption{Left panel: scale determination for three ensembles at $\beta=5.3$ with pion masses.
Right panel: comparison of the correlator $t^3G(t)$ computed on the lattice,
and its phenomenological determination from $e^+e^-$ data.
The normalization is such that $G(t) = (4\pi^2 t^3)^{-1}$ 
for non-interacting quarks.} 
\label{fig:ScaleDet}
\end{figure}

We are applying the scale setting method for the first time. We use
$N_{\rm f}=2$ ensembles of O($a$) improved Wilson fermions generated
as part of the CLS effort\footnote{{https://twiki.cern.ch/twiki/bin/view/CLS/WebIntro}}
 (see for instance~\cite{Francis:2013fzp} for
a more detailed description and references).  We choose the local
vector current at the source and the conserved vector current at the
sink, as in~\cite{Francis:2013fzp}.

Figure \ref{fig:ScaleDet} shows the lattice data for the function
$f(t)$, with $t$ rescaled in units of $\tau_1$. Results from three
ensembles at the same lattice spacing and for different pion masses
are shown. We observe a significant interval around $t=\tau_1$ where
the phenomenological curve and the lattice data are statistically
consistent, with little dependence on the pion mass observed. This is
an encouraging feature of the method. It implies that the result for
$\tau_1/a$ is insensitive to the precise choice (\ref{eq:325}) within
the statistical uncertainty.

\begin{figure} 
\centerline{\includegraphics[width=.5\textwidth]{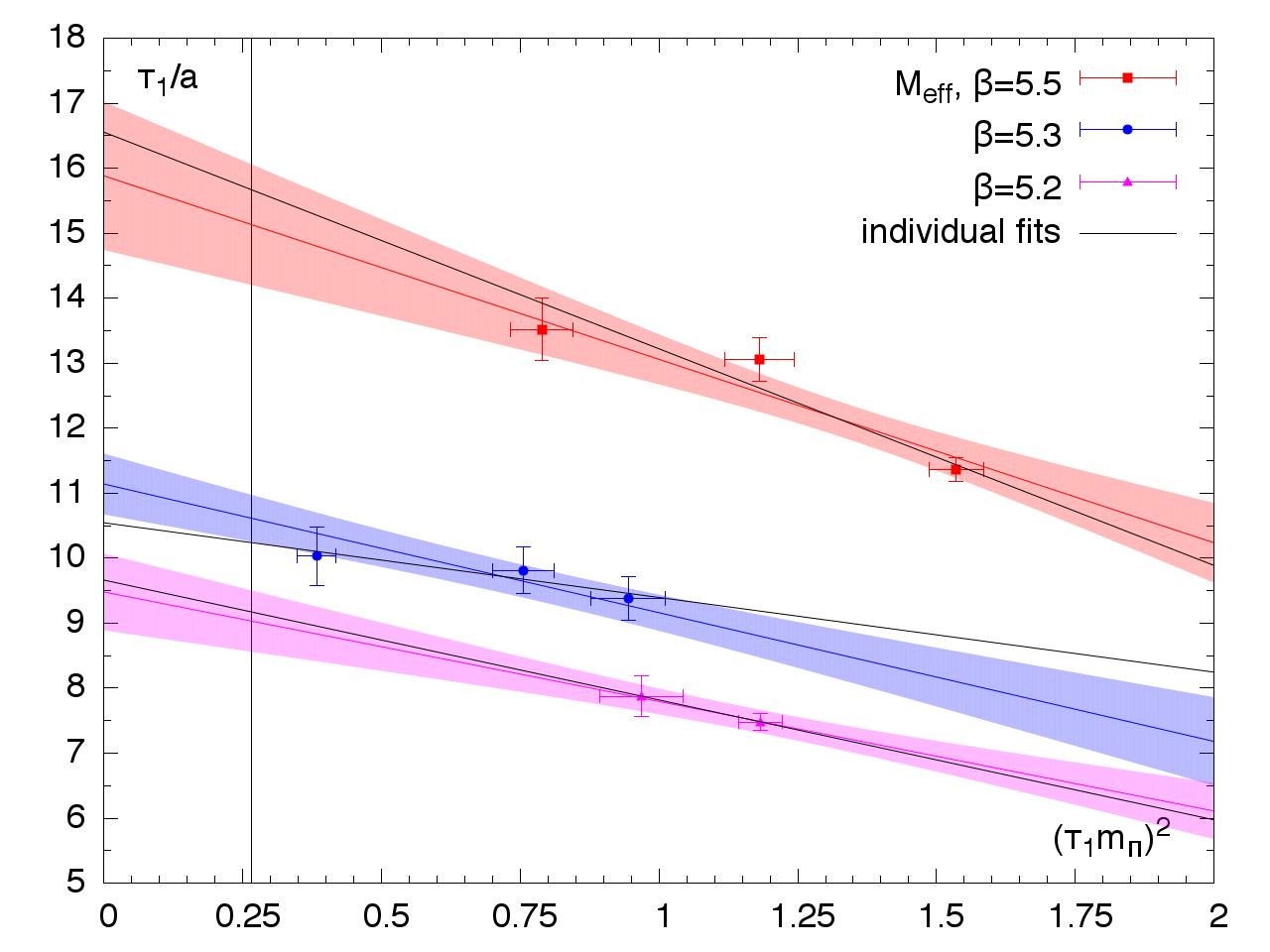} 
\includegraphics[width=.5\textwidth]{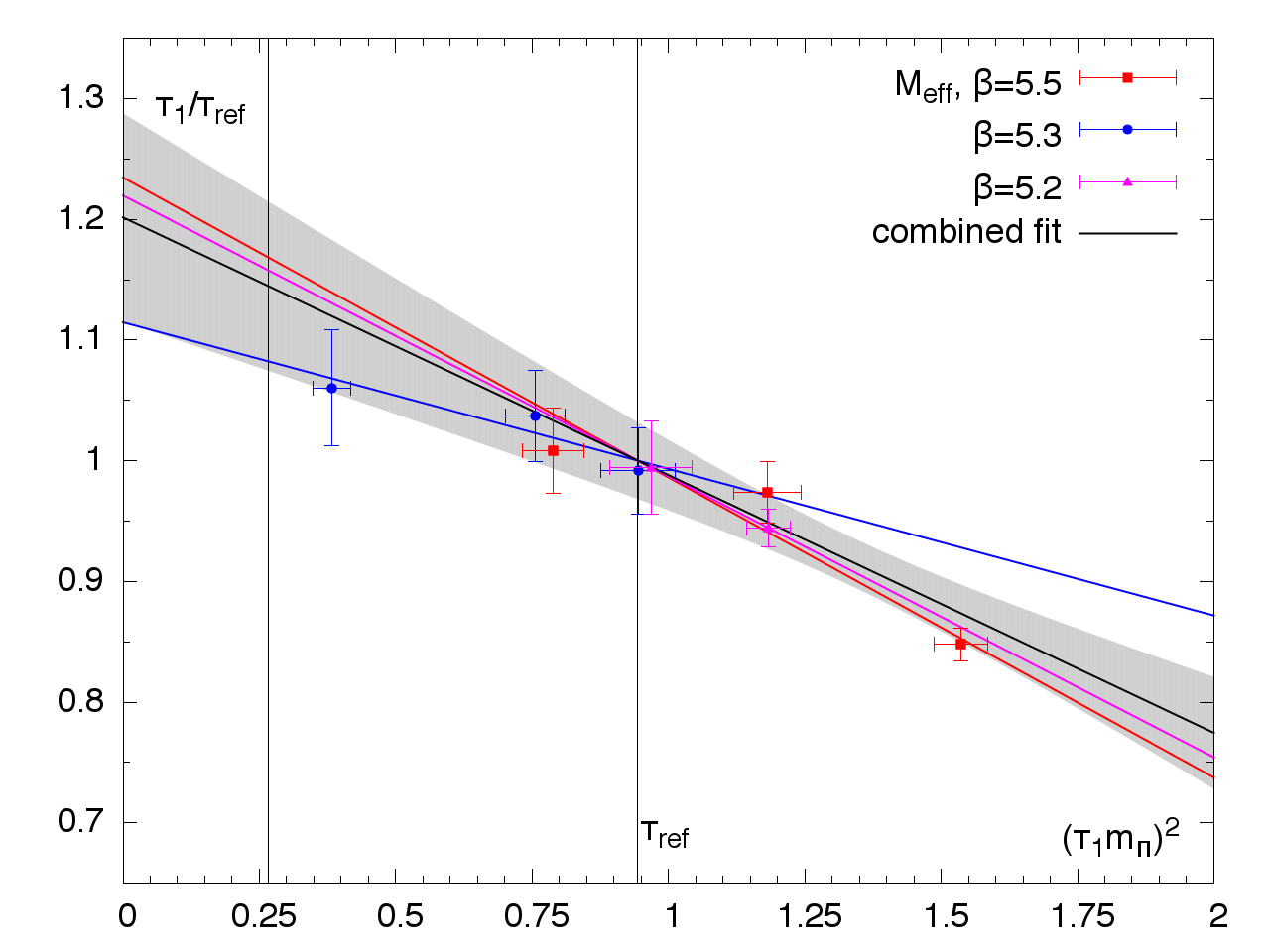} }
\caption{Left: chiral extrapolation of $\tau_1/a$ at three different values of $\beta$. 
Right: test of the cutoff effects on the quark mass dependence of $\tau_1$.} 
\label{fig:ChExtrapol}
\end{figure}


For the chiral extrapolation of $\tau_1/a$ viewed as a function of
$x\equiv(\tau_1 m_\pi)^2$, we adopt the point of view that we Taylor
expand this quantity around an intermediate value of $x$.  The
analysis then proceeds similarly as for $r_0$~\cite{Fritzsch:2012wq},
with the exception that we extrapolate $\tau_1/a$ to the physical
value of $x$ rather than $x=0$.  Given the number of data points we
have and their accuracy, we restrict ourselves to linear order in the
Taylor expansion. In the left panel of Fig.\ \ref{fig:ChExtrapol}, we
illustrate the extrapolation assuming that the pion mass dependence
has no cutoff effects. This assumption is tested in the right panel,
where the different $\tau_1$ values are interpolated to a common,
reference value of $\tau_1 m_\pi$, yielding $\tau_{1,\rm ref}$.  The
dependence of $\tau_1/\tau_{1,\rm ref}$ on $\tau_1 m_\pi$ should then
be universal.  Within the rather large uncertainties, this is what is
observed. We note that there is some evidence for a flattening of the
$m_\pi$ dependence at $\beta=5.3$, where we have access to a pion mass
below 200\,MeV.  Higher statistics are needed to establish this effect
with confidence.

The results for the lattice spacings, where the absence of cutoff
effects on the quark mass dependence of $a\tau_1$ has been assumed,
are given in Table \ref{tab:latspac}. Only a statistical error is
quoted here for these preliminary results. Our results appearing in
the first column are compared with those obtained previously via other
reference quantities, the Omega baryon mass~\cite{Capitani:2011fg} and
the kaon decay constant~\cite{Fritzsch:2012wq}. We note an overall
agreement with the latter results. There is a slight tension at
$\beta=5.3$ between our value of the lattice spacing and the value
obtained via the Omega mass.  However, unlike the other scale
determinations of Table \ref{tab:latspac}, the present one included
the G8 ensemble with a pion mass below 200\,MeV. As can be seen from
Fig.\ \ref{fig:ChExtrapol}, without this ensemble, the lattice spacing
would come out slightly smaller.

We can turn to the comparison of the correlation function $G(t)$
itself to its phenomenological determination from $e^+e^-\to {\rm
  hadrons}$ data (right panel of \fig\ref{fig:ScaleDet}).  Apart from
the expected large cutoff effects at short distance, we observe that
the fall-off of the lattice correlators is faster for the larger pion
masses, as one would expect.  On the G8 ensemble at $m_\pi=190$\,MeV,
the fall-off appears to be slower, but the statistical uncertainty
becomes large beyond 1.2\,fm.

\begin{table}[t]
\centerline{
\begin{tabular}{c@{~~~}c@{~~~}c@{~~~}c} 
\hline
$\beta$ & $a/ {\rm fm}$ from $\tau_1$ &  $a/ {\rm fm}$ from $m_\Omega$ \cite{Capitani:2011fg} & $a/ {\rm fm}$ from $F_K$ \cite{Fritzsch:2012wq}\\
\hline
5.5 & 0.048(3)   & 0.050(2)(2)  & 0.0486(4)(5)  \\
5.3 & 0.0685(23)  & 0.063(2)(2)  & 0.0658(7)(7)  \\
5.2 & 0.081(4)    & 0.079(3)(2)   & 0.0755(9)(7)  \\
\hline 
\end{tabular}}
\caption{Comparison of lattice scale determination between our (preliminary) results 
and results based on the $\Omega$ baryon mass~\cite{Capitani:2011fg}
and the kaon decay constant $F_K$~\cite{Fritzsch:2012wq}.}
\label{tab:latspac}
\end{table}

\section{Systematic effects}

\begin{figure} 
\centerline{\includegraphics[width=.5\textwidth]{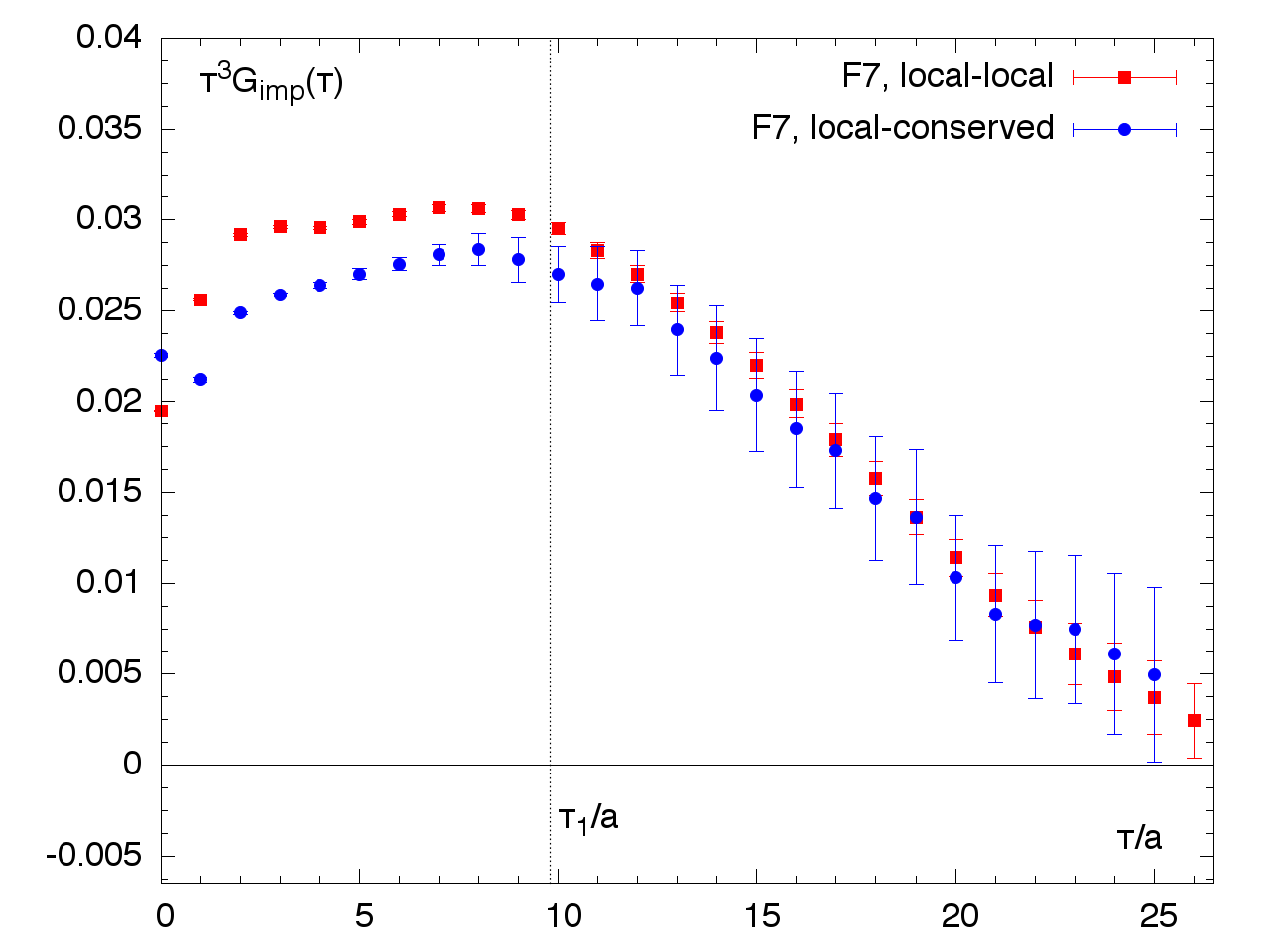}
\includegraphics[width=.5\textwidth]{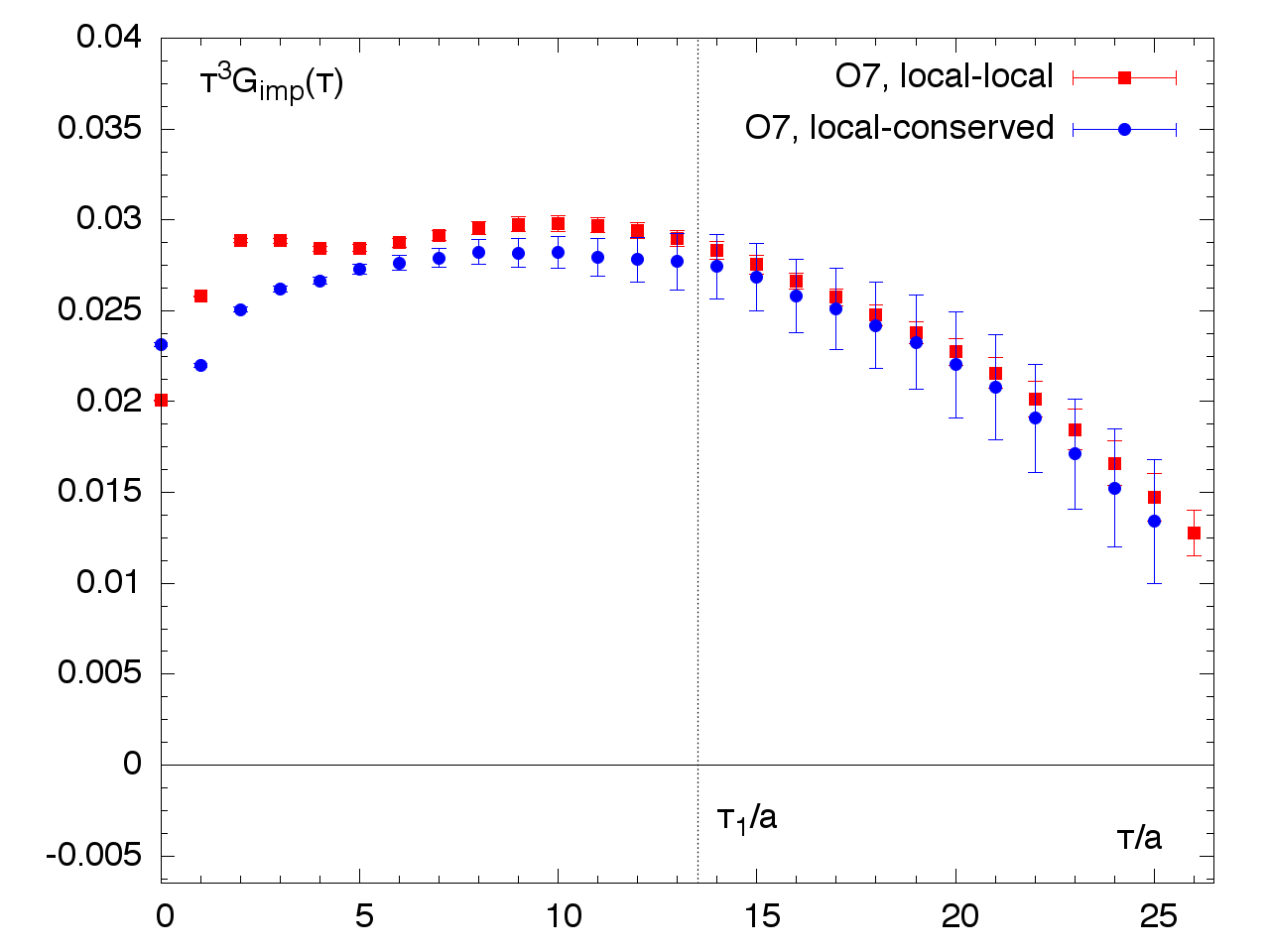}
}
\caption{A comparison of two discretizations of the vector correlator at $m_\pi\approx 270{\rm \,MeV}$. 
Left: $\beta=5.3$, corresponding to 
$a\simeq 0.068{\rm \,fm}$. Right: $\beta=5.5$, correspondings to $a\simeq 0.048{\rm \,fm}$.} 
\label{fig:CutOff}
\end{figure}

\noindent\emph{Cutoff effects:} At a pion mass of approximately 270\,MeV, we present 
a comparison of two vector correlators, one computed using the lattice vector current $\bar\psi\gamma_\mu\psi$
at both source and sink and the other one using instead the conserved vector current at the sink
(i.e.\ the discretization we used for the scale determination).
Clearly, we expect any difference to be reduced when we go to a smaller lattice
spacing. A comparison is shown in \fig\ref{fig:CutOff}. The
improvement terms in the vector current have not been included
($b_V=c_V=0$ in the notation of~\cite{Sint:1997jx}). However the correlators
are tree-level improved, i.e.\ they have been divided by the free lattice 
correlator (see appendix B of \cite{Brandt:2013faa}) and multiplied by the free continuum correlator
(tree-level improvement was also applied in the right panel of \fig\ref{fig:ScaleDet}).
The cutoff effects are observed to be large for $t<0.5{\rm \,fm}$ but are 
reduced, as expected, when going from $\beta=5.3$ to $\beta=5.5$.
It would be interesting to see whether the improvement term proportional to $c_V$
would reduce the size of the cutoff effects.

\noindent\emph{Finite-size effects} on the vector correlator are expected to be
exponential at fixed time separation $t$~\cite{Francis:2013fzp},
however the convergence to the infinite-volume limit is non-homogeneous
in $t$. At large times, when the correlator is dominated by the few lowest-lying states,
the dominant finite-size effects can be understood using L\"uscher's finite-volume 
formalism~\cite{Luscher:1991cf,Meyer:2011um}.
At the intermediate distance of about 0.7\,fm, we expect to be in the regime where 
the finite-size effects are falling off exponentially in the box size. Obviously it would be 
desirable to check this explicitly.

\section{Discussion}

We have implemented a method to determine the energy scale in lattice QCD.
The results are compatible with previous determinations on the same $N_{\rm f}=2$ ensembles.
It is desirable at this point to improve the statistical accuracy, perhaps 
by using noise-reduction techniques. Also,  O($a$) improvement should be implemented
at the level of the operator.

The method  presented here straightforwardly
generalizes to 2+1 and 2+1+1 flavors, where the main ambiguity of
separating the different flavors disappears. 
Our scale determination method is relatively insensitive to the quark masses, see \fig\ref{fig:ChExtrapol}.
The Omega mass, for instance, depends quite
strongly on the strange quarks mass, therefore the uncertainty on the
scale determined via this quantity must take into account the
uncertainty in the strange quark mass setting.

For the isovector channel considered here, it is also attractive to
use $\tau$ decay data on the phenomenology side~\cite{Golterman:2013vca}. 
We intend to investigate this possibility in a longer forthcoming publication.

While it is difficult, at present, for lattice calculations to compete
with the high precision of $e^+e^-$ data for the determination of
$(g-2)_\mu$, flavor combinations in the vector and axial-vector
channel that are not directly accessible experimentally should
eventually be calculated on the lattice. In particular, this could
have an impact on the precision determination of the running of the
weak mixing angle.


{We are grateful 
to our colleagues within CLS for sharing the lattice ensembles used in this study.  
We also thank the participants of the MITP scientific program 
\emph{Low-Energy Precision Physics} (Mainz, 23~Sep.\ -- 11~Oct.\ 2013) for helpful discussions.
The correlation functions were computed 
on the dedicated QCD platform ``Wilson'' at the Institute for Nuclear Physics,
University of Mainz. 
This work was supported by the \emph{Center for Computational Sciences}
as part of the Rhineland-Palatinate Research Initiative.}

\bibliographystyle{JHEP}
\bibliography{/Users/harvey/BIBLIO/viscobib}

\end{document}